\documentclass[english]{elsarticle}
\usepackage[T1]{fontenc}
\usepackage[latin9]{inputenc}
\usepackage{textcomp}
\usepackage{amstext}
\usepackage{amssymb}
\usepackage{graphicx}
\usepackage{esint}
\usepackage{natbib}
\graphicspath{{./}}

\makeatletter
\journal{European Economic Review}

\makeatother

\usepackage{babel}
\begin{document}

\begin{frontmatter}{}

\title{\textbf{Controlling volatility of wind-solar power}}

\author{Hans Lustfeld}

\ead{h.lustfeld@fz-juelich.de}

\address{Peter Grünberg Institut (PGI-1), Forschungszentrum Jülich 52425 Jülich,
Germany}
\begin{abstract}

The main advantage of wind and solar power plants is the power production
free of $CO_{2}$. Their main disadvantage is the volatility of the
generated power. According to the estimates of H.-W. Sinn\citep{key-1},
suppressing this volatility requires pumped-storage plants with a
huge capacity, several orders of magnitude larger than the present
available capacity in Germany\citep{key-2}. Sinn concluded that wind-solar
power can be used only together with conventional power plants as
backups. However, based on German power data\citep{key-3} of 2019
we show that the required storage capacity can significantly be reduced,
provided i) a surplus of wind-solar power plants is supplied, ii)
smart meters are installed, iii) partly a different kind of wind turbines
and solar panels are used in Germany. Our calculations suggest that
all the electric energy, presently produced in Germany, can be obtained
from wind-solar power alone. And our results let us predict that wind-solar
power can be used to produce in addition the energy for transportation,
warm water, space heating and in part for process heating, meaning
an increase of the present electric energy production by a factor
of about 5\citep{key-1}. Of course, to put such a prediction on firm
ground the present calculations have to be confirmed for a period
of many years. And it should be kept in mind, that in any case a huge
number of wind turbines and solar panels is required.
\end{abstract}

\end{frontmatter}{}

\section{Introduction}

Apart from nuclear power and hydropower (power from biomass and waste
could be mentioned, too), the conventional electric power production
by gas and fossil fuel power stations generates $CO_{2}$ as a byproduct.
Nuclear power plants do not have this problem, but they have other
disadvantages, in particular production of radioactive waste. All
these problems do not appear when electric power is produced by solar
panels and wind turbines alone. Nevertheless, wind-solar power has
serious disadvantages too, the most serious one being the volatile
energy production: Weather conditions change rapidly and as a consequence
energy production of wind-solar power fluctuates considerably. How
serious the consequences are, depends on two factors: i) the strength
of the volatility, ii) the volatility that a consumer can tolerate
- cf. the key phrase \textquotedblleft new thinking\textquotedblright \citep{key-4}.

The problem of volatility is always a consequence of the mismatch
between electric power production and electric power consumption.
If we just consider passive storage devices, like pumped-storage plants,
for example, the strength of volatility can be determined in the following
way. We divide the wind-solar energy production $P_{v}$ and the energy
demand $P_{d}$ into two parts: the average parts $P_{va}$ and $P_{da}$
being constant over the year and the fluctuating parts $P_{vf}$ and
$P_{df}$. The power $P_{sf}$ streaming into and out of storage devices
has a fluctuation part only since the devices are passive. The average
of each fluctuating part is zero. Therefore we get the simple equations
\begin{equation}
P_{va}=P_{da}\label{eq:averageEQ}
\end{equation}
and therefore,
\begin{equation}
\begin{array}{ccc}
P_{sf} & = & P_{vf}-P_{df}\end{array}\label{eq:SpeicherGl}
\end{equation}

Integrating these powers yields the energies
\begin{equation}
\begin{array}{ccc}
E_{v(d)}(t) & = & E_{va(da)}(t)+E_{vf(df)}(t)\\
E_{va(da)}(t) & = & P_{va(da)}\cdot t\\
E_{vf(df)}(t) & = & \intop_{0}^{t}P_{vf(df)}dt\\
E_{sf} & = & E_{vf}-E_{df}
\end{array}\label{eq:LeistungIntegrale}
\end{equation}
The required storage $E_{sfmax}$ is obtained from the expression
\begin{equation}
E_{sfmax}=\max_{t}\{E_{sf}(t)\}-\min\{E_{sf}(t)\}\label{eq:Speichergroesse}
\end{equation}

\begin{figure}
	\includegraphics[scale=0.42]{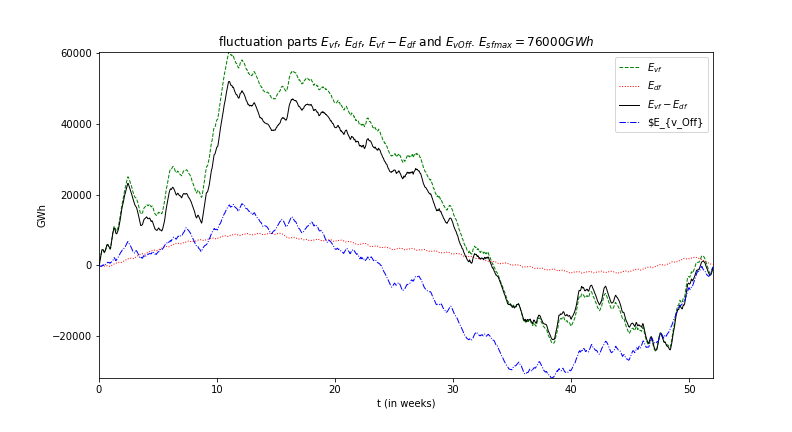}
	\caption{Fluctuation parts and storage: (Green) dashed line: Fluctuating part
$E_{vf}$ of (scaled) volatile energy $E_{v}$. (Red) dotted line:
Fluctuation part $E_{df}$ of energy demand $E_{d}$.(Black) solid
line: (Scaled) fluctuating part $E_{vf}$of wind-solar energy minus
fluctuating part $E_{df}$ of demand $E_{d}$. From the difference
between max and min the required storage can be read off. Note the
dominance of $E_{vf}$. For comparison the fluctuation part $E_{vfOff}$
of the (scaled) offshore energy $E_{vOff}$is shown. Here the scaling
factor is $56.4/2.76$\label{fig:fig01}}
\end{figure}

In order to obtain these functions for 2019, we have used the data
of ref\citep{key-3} which contain power measurements every $15$
minutes {[}in MW{]} for electric load and the volatile power, consisting
of: offshore and onshore windpower as well as solar power. The result
is obtained directly for the demand (load): $P_{da}=56.4$GW and the
fluctuation part $E_{df}$ is plotted in Fig.\ref{fig:fig01}. The
correct values for $P_{va}$ and $P_{vf}$ require a moment of consideration.
The measurement data lead to the value $\hat{P}_{va}=18.9$GW.

 Since we intend to satisfy the average electric energy demand by
wind-solar power alone, we need an average electric power production
of at least $P_{va}=P_{da}=\text{56.4}$GW as well. To manage this
problem, we assume that the distribution of solar cells and wind turbines
is already at its optimum in Germany. In that case we can easily estimate
the situation where all average electric energy is delivered by wind-solar
power: We just have to multiply the average wind-solar power and its
fluctuation by a scaling factor\citep{key-1}. Clearly the distribution
of wind-solar power is not at its optimum in reality. Therefore, the
scaling is an approximation. However, understanding the trends is
the aim of this paper and for that purpose the scaling will be a sufficiently
good approximation. As the distribution becomes increasingly optimized,
such an approximation will steadily improve. Now taking as natural
scaling factor $56.4/18.9$ for wind-solar power we get
\begin{equation}
\begin{array}{ccc}
P_{va} & = & \hat{P}_{va}\cdot\frac{56.4}{18.9}=P_{da}\\
\\
P_{vf} & = & \hat{P}_{vf}\cdot\frac{56.4}{18.9}
\end{array}\label{eq:scaling}
\end{equation}

where $\hat{P}_{va}$ and $\hat{P}_{vf}$ are obtained from the measurement
data. In the same way we get the scaled integrated quantities $E_{v},$$E_{va}$
and $E_{vf}$ after scaling the original mathematical expressions
with the same factor. Together with the scaling factor $56.4/18.9$
those expressions will be used throughout this paper. Of course, wind-solar
and demand data differ from year to year, but, considered over the
year 2019, the typical properties should be similar enough to reveal
the relevant trends. For a detailed technical analysis, the observation
period of one year does, of course, not suffice. Instead, observation periods
over many years are necessary. This is possible, but beyond the scope
of the present paper.

As a first application we calculate the storage capacity necessary
to fulfill eq.{[}\ref{eq:SpeicherGl}{]} and we find: the required
electric storage removing volatility is $75.9TWh$, cf. eq.\ref{eq:Speichergroesse}
and Fig.\ref{fig:fig01}.

This result proves that volatility is a serious problem. Indeed volatility
led Sinn to the conclusion that energy production resting essentially
on wind-solar power alone will take us into an economic nirvana\citep{key-5}.
One could argue that even a total storage capacity of about $84$TWh\citep{key-1},
\citep{key-2} is in principle feasible by transforming the huge Norwegian
hydro dams into pumped-storage plants. However, two facts are obvious:
i) The present electric power production has to be multiplied by a
factor of about $5$ \citep{key-1}, if all transportation, warm water,
space heating and a considerable percentage of process heating are
switched to electric power as well. With the configuration presented
here so far, this is impossible. ii) Even if we do not consider transportation,
warm water, space heating, and process heating, the pump storage capacity
requirements would be so enormous that an export of this wind-solar
scheme to many other nations would be out of the question - a bitter
disadvantage when Germany wants to be a forerunner.

How can we successfully proceed? Of course we need a kind of active
buffering. Sinn \citep{key-1} suggests various combinations between
wind-solar power and conventional power plants. This works for the
present needs but eliminates the advantages of using wind-solar power
alone. Furthermore, this scheme becomes questionable if the needs
of electric power are five times greater than now. And that will happen
if energy production of Germany will be switched to electric power
generation as far as possible.

In this paper we discard such schemes. Instead we suggest a combination
of three methods: First, creating a surplus of wind-solar power. Second,
applying smart meters\citep{key-4}. Both methods are described in
section II. We show that through these methods the storage capacity
requirements are reduced by a factor of about 50 or even become marginal.
Third, applying new criteria for optimizing the efficiency of wind
turbines, solar cells and their distribution across the country. We
show in section III that through these additional methods the smart
meters need distinctively less flexibility. We think that these results
give us the justification for extrapolating to the case where in addition
to the present electrical energy production all energy for the total
transport, warm water, space heating and and a considerable percentage
of process heating is exclusively obtained from wind-solar power.
This is discussed in section IV. Our conclusions are presented at
the end of the paper.

\section{Surplus of wind-solar power and smart meters}

In this section we discuss the situation, in which the generation
of electric power is taken over by wind-solar power alone.

In contrast to simple passive buffers like pumped-storage plants with
all their capacity limitations active buffers like power plants can
satisfy the demand necessary for guaranteeing a safe power delivery.
To avoid $CO_{2}$ production we choose as active buffers wind-solar
power itself. Assuming\citep{key-1} as above an already optimum distribution
of wind-solar power devices across the nation, the surplus of wind-solar
power can be expressed again by a scaling factor , the strength $\alpha$,
and we get for the wind-solar energy 
\[
E_{v}(t)\longrightarrow(1+\alpha)E_{v}(t),\alpha=const
\]
The price to be paid for this scheme is a reduced efficiency. This
is all the more the case since at times of low wind-solar power the
surplus of wind-solar power is reduced as well enforcing a larger
$\alpha$ value than expected from the average gain in power. To keep
$\alpha$ within reasonable limits we need the concept\citep{key-6}
and on a large scale the application\citep{key-7} of smart meters\citep{key-4}.
Such devices control the electrical consumption very effectively by
setting higher consumption prices if less power is available and lower
prices if there is a surplus of power. Thus smart meters act like
passive buffering devices by moving the peaks of electric consumption
to the peaks of wind-solar power. We note in passing that moving electric
consumption on an hourly basis or less has the same effect as smoothing,
cf. ref. \citep{key-1}, section 4.

Of course a detailed simulation of smart meters is intricate. However,
we think that the following simulation of smart meters reflects the
principle effects satisfactorily: $E_{d}(t)$ is the energy of electric
consumption as function of time. Now, if wind-solar production is
not sufficient, it produces energy corresponding to a demand $E_{d}(t')<E_{d}(t)$.
The smart meters have now the task, by increasing prices for $1kWh$
to enforce a reduced demand, namely $E_{d}(t')$. Clearly that is
always possible - due to exorbitant prices, if necessary. But to avoid
such uneconomic incidents we introduce the delay time $\tau\geqslant0$
and require $t'>t-\tau$. The analogous happens for a surplus of wind-solar
power: If there is an excess production of electric energy, corresponding
to $E_{d}(t')>E_{d}(t)$, smart meters charge low prices and $E_{d}$
is again not that of time t but that of $t'$ and $t'<t+\tau$. This
means the strict requirement $t'=t$ is replaced by 
\begin{equation}
E_{d}(t-\tau)<E_{d}(t')<E_{d}(t+\tau),\,\mid t'-t\mid\leq\tau\label{eq:tauInterval}
\end{equation}
As is the case in the applications of real smart meters, the electric
power consumption becomes more flexible in this simulation. The produced
energy till time t need not be $E_{d}(t)$ exactly but staying in
the interval of eq.(\ref{eq:tauInterval}) suffices. This flexibility saves
storage capacity as can be seen from an extreme (hypothetical) case:
If $\tau$ becomes sufficiently big the required storage capacity
approaches zero. Of course such large values of $\tau$ are unrealistic.
However, values of hours or even days can be acceptable. We assume
that a limit $\tau\leq1$day is a very reasonable one. And note, this
simulation of smart meters fulfills the criterion that - as in real
smart meters - consumption of energy is only moved, no energy is generated
or lost.

To avoid large $\tau$ values, finite $\alpha$-values and possibly
electric buffering devices are still needed. We find out the relation
between delay time $\tau$, capacity $E_{sfmax}$, and strength $\alpha$
in the following manner:

We fix $\alpha$ and $\tau$, proceed in time steps $\Delta t=15$min
and define $E_{d}(t')=\breve{E_{d}}$. At the beginning we set $\breve{E}=0$
and $E_{sf}=0$. Now at time $t+\Delta t$ we get
\[
\begin{array}{ccc}
\delta E_{v} & = & (1+\alpha)P_{v}(t)\cdot\Delta t\end{array}
\]
 Now three cases have to be distinguished (we set $s=t+\Delta t$):

i) $E_{d}(s-\tau)\leq\breve{E}_{d}+\delta E_{v}\leq E_{d}(s+\tau)$,
then $\breve{E}_{d}\rightarrow\breve{E}_{d}+\delta E_{v}$.

ii) $\breve{E}_{d}+\delta E_{v}>E_{d}(s+\tau)$, then $\breve{E}_{d}\rightarrow E_{d}(s+\tau)$
and $E_{sf}\rightarrow E_{sf}+\breve{E}_{d}+\delta E_{v}-E_{d}(s+\tau)$

iii) $\breve{E}_{d}+\delta E_{v}<E_{d}(s-\tau)$, then $\breve{E}_{d}\rightarrow E_{d}(s-\tau)$
and $E_{sf}\rightarrow E_{sf}+\breve{E}_{d}+\delta E_{v}-E_{d}(s-\tau)$

Furthermore, to save storage, we enforce $E_{sf}\leq0$ by replacing
a positive $E_{sf}$ with zero. Thus a positive $E_{sf}$ becomes
a kind of 'wasted' energy that has to be removed somehow (see below).
At the end of the calculation $E_{sfmax}$ is given by 
\[
E_{sfmax}=\max_{t}\{-E_{sf}(t)\}
\]

Repeating this procedure for various $\tau$ values we get the function\\
$E_{sfmax}(\tau,\alpha)$ and from that function the inverse function
$\tau(E_{sfmax},\alpha)$.
\begin{figure}
\includegraphics[scale=0.41]{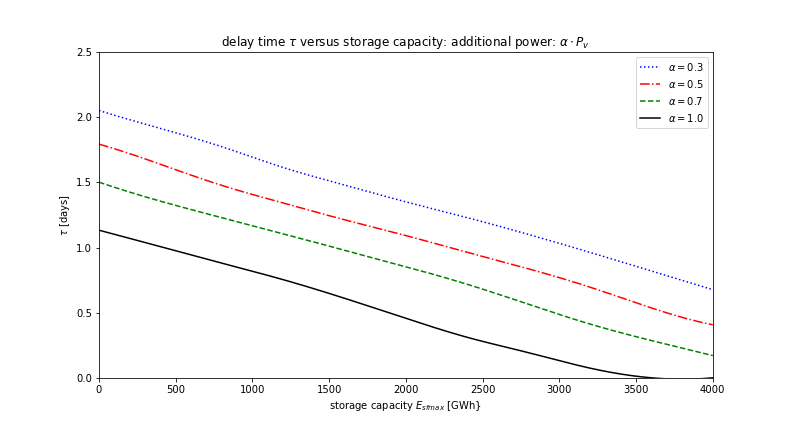}
\caption{Delay time $\tau${[}days{]} versus storage capacity $E_{sfmax}[GWh]$
for various values of the strength $\alpha$. The curves demonstrate
that a decrease of the storage capacity can indeed be compensated
by an increase of $\alpha$. E.g. $\tau=1$day can be obtained by
setting $\alpha=0.3$ and storage capacity of $3100$ GWh or $\alpha=0.5$
and storage capacity of 2400 GWh or $\alpha=0.7$ and storage capacity
of $1500$ GWh or $\alpha=1.0$ and storage capacity of $400$ GWh.\label{fig:delayTime}}
\end{figure}
 
\begin{figure}
\includegraphics[scale=0.42]{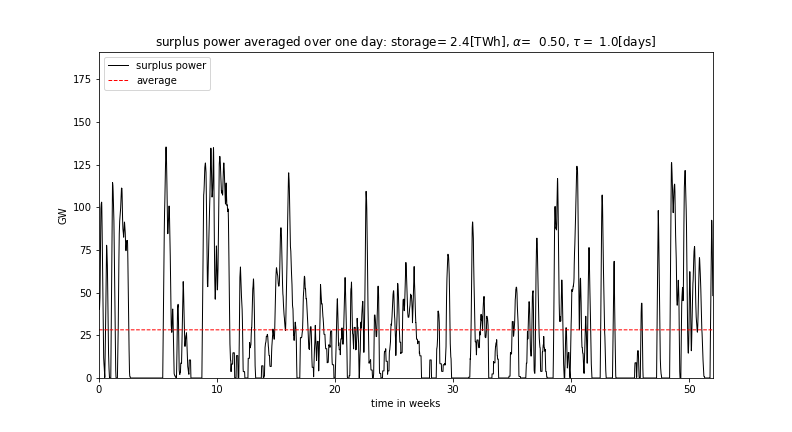}
\caption{'Wasted' power for storage capacity $2400GWh$, $\alpha=0.5$, $\tau=1.0$
days. This power is generated, if the surplus wind-solar devices work
all the time with maximum power available. (Black) solid line: 'Wasted'
power averaged over $24h$. (Red) dashed line: 'Wasted' power averaged
over one year. The averaged value is $\alpha\cdot$ $56.4$GW. \label{fig:SurplusFluctuation}}
\end{figure}

Results of our calculations are shown in Fig.\ref{fig:delayTime}.
We have plotted $\tau(E_{sfmax},\alpha)$ for various fixed $\alpha$
values. The results contain the important conclusion that storage
capacity can be replaced by a surplus of wind-solar power. This phenomenon
opens a wide field of possibilities: If storage capacity does not
represent a problem $\alpha=0.3$ might be sufficient. On the other
hand storage capacity becomes rather uncritical for $\alpha\geqq0.5$.
And for $\alpha=1$ the storage capacity is no longer a problem.

The arising 'wasted' energy need not be small at all. In fact, if
all possible energy is generated the average power amounts to $(1+\alpha)\cdot56.4GW$
and thus the average 'wasted' power to $\alpha\cdot56.4GW.$ To get
rid of it directly is one possibility. This can easily be achieved
by reducing the wind-solar power, as soon as 'wasted' energy begins
to build up. The advantage of this procedure would be a strain imposed
on the electricity network that would not essentially be higher than
for $\alpha=0$.

An alternative would be exploiting this 'wasted' power for processes,
e.g. for electrolytic and chemical processes. However one has to keep
in mind that the surplus power is really extremely volatile as can
be seen from Fig.\ref{fig:SurplusFluctuation} for $\alpha=0.5$.
Apart from high peaks there are - more important - periods, even weeks,
where no 'wasted' power is available.

The absolute costs per $kWh$ depend on assumptions, of how prices
will develop in the future and which indirect costs have to be included
in the calculation and which not. In fact, the estimates fluctuate
strongly\citep{key-1}\citep{key-8}\citep{key-9}. However, the relative
increase of running costs per $kWh$ due to the 'wasted' power can
be assessed: For a small contribution of wind-solar power - so small,
that peaks do not overshoot consumption - suppose running costs in
the average to be $ws_{small}]${[}€/kWh{]}. However, we do not deal
with a small contribution. Instead an average demand of $56.4GW$
has to be satisfied. Applying the surplus power approach fulfilling
this demand requires an average production of $(1+\alpha)\cdot56.4GW$.
Therefore, the increase of the running costs is $ws_{small}\rightarrow ws_{small}\cdot(1+\alpha)$
and the relative increase is given by $\alpha$.

\section{Low energy production and offshore wind turbines}

At first sight it may seem obvious that wind-solar power should have
its nominal power at high winds, at high sun radiation and moreover
in regions with high winds and high sun-radiation, respectively. But
the surplus wind-solar power becomes important if the wind-solar power
production is weak and therefore weak-wind turbines and solar cells
with good low-light performance will be essential for good additional
power production. Weak-wind turbines having blades enlarged by a factor\citep{key-10}
$\sqrt{\beta}$, greater height\citep{key-11} and thus higher wind
speed enlarged by a factor\citep{key-12} $\left(\gamma\right)^{1/3}$,
provide an increase of power generation by a factor of $\beta\cdot\gamma$.
We choose $\beta\cdot\gamma=2.$ This doubles the power production
in the low-wind regime, $P_{l}=2\cdot P$. In the high wind regime,
however, the power production saturates since these turbines have
a reduced nominal power\citep{key-12} $P_{nom}$. This justifies
the ansatz

\[
P_{l}(t)=P_{nom}\cdot\tanh\left(\beta\cdot\gamma\cdot P(t)/P_{nom}\right),\,\beta\cdot\gamma=2
\]

Weak-light performance of solar cells\citep{key-13} depends on the
material used\citep{key-14}. Mono-crystalline PV modules\citep{key-15},
multi junction\citep{key-16} with selected band gaps and in the future
the new generations of DSSCs\citep{key-17}\citep{key-18} may have
good weak light performance. And we assume that with good weak light
performance the generated power can increase - as for the wind power
- by a factor of $2$ also in the weak-light regime. (This approximation
may be crude but less important as well, since the capacity factor
of solar cells is dismal, at least in Germany\citep{key-19}, $10-13$
\%, and therefore the dominant power generation will be that of wind
turbines). So we get for the total surplus power (here defined as
$P_{low}$) the ansatz
\begin{equation}
P_{low}(t)=P_{va}\cdot\tanh\left(2P_{v}(t)/P_{va}\right)\label{eq:lowE}
\end{equation}

and 
\[
\begin{array}{ccc}
\delta E_{v} & = & \left(P_{v}(t)+\alpha P_{low}(t)\right)\cdot\Delta t\end{array}
\]

\begin{figure}
\includegraphics[scale=0.21]{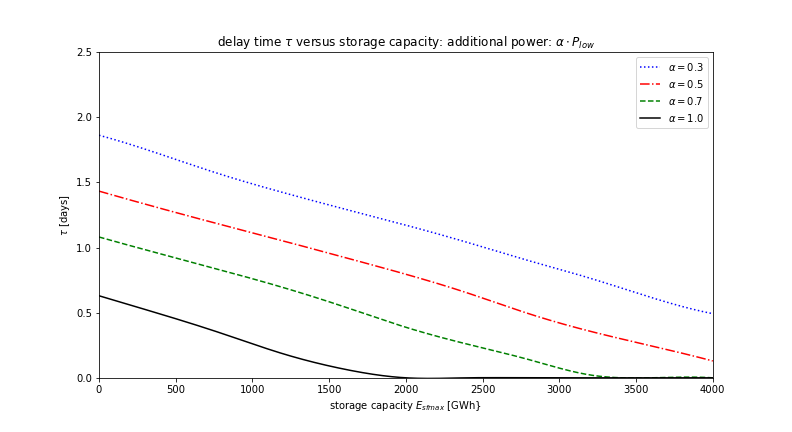}
\includegraphics[scale=0.21]{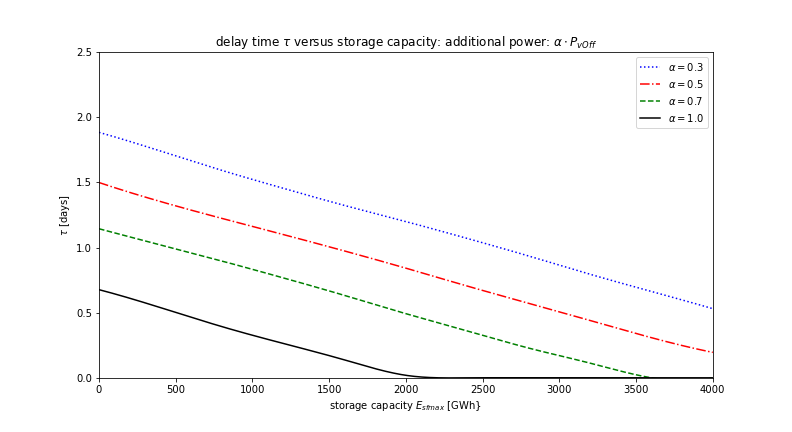}
\caption{Delay time $\tau${[}days{]} versus storage capacity $E_{sfmax}${[}GWh{]}
for various values of the strength $\alpha$. Left: The surplus power
$\alpha P_{low}$has enhanced performance at low wind and low radiation,
cf. eq.{[}\ref{eq:lowE}{]} leading to distinctively shorter delay
times $\tau$. Right: The surplus power $\alpha P_{vOff}$ is delivered
by offshore wind turbines. Their averaged contribution to the electric
power generation is still small, only $2.8$GW in $2019$. However,
a continuous expansion is intended, therefore we have scaled the present
offshore power generation by a factor of $56.4/2.8$. So for $\alpha=0.5$
the required offshore power has to exceed the value of 2019 by a factor
of 10. According to our calculations such an expansion would lead
to promising results: Again distinctively shorter delay times $\tau$.\label{fig:Delay-timelow}}
\end{figure}

To demonstrate the importance of low energy production, we have selected
a low nominal power $P_{nom}$ for $P_{low}$: $P_{nom}=P_{va}.$
And the average wind-solar power is about five times less than the
normal nominal power.

The calculations correspond to those of the previous section. Results
are presented in Fig. \ref{fig:Delay-timelow}. The distinctly better
outcome for the delay times $\tau$ is obvious in spite of the low
nominal power $P_{va}$ of $P_{low}$. This proves the importance
of good performance in weak wind and low light situations.

We found a similar improvement of the results when using offshore
wind turbines. We repeated the calculations, only replacing the weak
wind power by scaled offshore wind power. With a value of about $20$
the scaling factor guaranteeing $P_{va_{Off}}=P_{da}$ is rather large
since at present the offshore power amounts to $5$\% of the electric
load (average) only, cf. Fig.\ref{fig:Delay-timelow}..

\section{Addition of all possible electric energy in Germany}

In the two preceding sections the possibility of applying solar-wind
power without excessive use of storage devices has been demonstrated.
However, we had only discussed the case of replacing the present electric
energy production by wind-solar power. But this amounts to about $20\%\thickapprox60GW$\citep{key-1}
over the year, whereas the total energy production amounts to $\thickapprox280GW$,
averaged over the year). $80\%$ consists of energy production for
transport on a fossile basis, warm water, space heating and process
heating. Converting this energy production into electric power should
be possible not completely but to a large extent.

Therefore, the question is inescapable: Can all this electric power
be generated by wind-solar power alone. I think a precise answer to
this question presumes inspecting all the fluctuation data of wind
and radiation across the various parts of Germany over many years.
This is possible but beyond the scope of the present paper.

On the other hand, when we look at the required electric storage results
for the various examples in the introduction we conclude that compared
to the less drastic volatility of electric consumption the volatility
of wind-solar power is the dominant part, cf. red dotted curve in
Fig.\ref{fig:fig01}. But this part can be estimated by simple scaling
as in section II, leading to a scale factor of $5$. This would mean
that the plots in Fig.\ref{fig:delayTime} and Fig. \ref{fig:Delay-timelow}
remain the same, only the storage values on the horizontal axis had
to be multiplied by a factor of $5$. Since the storage values are
not particularly critical, the effect of scaling would require a shift
$\alpha\rightarrow.6$ or $\alpha\rightarrow1$, which in our view
seems to be a tolerable change.

Furthermore we can argue that in spite of its enormous volatility
at least part of the 'wasted' power can be used for chemical, in particular
electrolytic processes, by which artificial fuel can be produced,
e.g. for airplanes. This again would reduce the required wind-solar
energy and thus the scaling factor.

It should also be pointed out that in any case the number of required
wind turbines is enormous. After conversion into electric power an
average power of $\thickapprox250GW[1]$\citep{key-1} has to be generated
in Germany. Assuming i) a capacity factor of $25$\% for wind turbines{[}18{]})
and ii) a $1/3$ contribution of solar power (probably a bit less\citep{key-20}),
a simple calculation leads to a required nominal wind power of $650000MW$.
This means $430000$ \{110000\} wind turbines of the $1.5MW$ \{6MW\}
type (height 120m \{200m\}) are needed to produce the average power.
But this is not enough. In our approach this number has to be increased
by $60$-100\% to control volatility.

In view of these large numbers manufacture of wind turbines in mass
production should be possible, reducing the cost of wind-solar power.

\section{Conclusions}

Is it possible to switch all present electric energy production of
Germany to wind-solar power? This question has been answered in the
negative by Sinn\citep{key-5}. The main topic of this paper is to
show that Sinn's judgement is too pessimistic. When choosing a different
ansatz, the results become different: We suggest marginalizing the
strong volatility of wind-solar power by i) adding a substantial surplus
of wind-solar power ii) installing smart meters\citep{key-4},\citep{key-6}
iii) selecting different kinds of wind turbines, solar devices and
switching to a good deal to offshore wind turbines.

The results of our ansatz are encouraging: The electric storage needed
is reduced by more than $90\%$, even nearly $100$\% should be possible.

The prize to be paid would be a $50$\% - $100$\% surplus of wind-solar
power devices compared to the situation in which only the averaged
wind-solar power production matches the averaged power consumption.

Our precise data\citep{key-3} extend over the year 2019, a period
sufficient to show that wind-solar power can be promising. Unambiguous
results will be confirmed when weather conditions in Germany are carefully
analyzed over a period of many years - but this is beyond the scope
of the present paper. However, based on the present data, measured
every 15 minutes in 2019, our approach avoiding excessive passive
storage leads to the following conclusions: First, this approach is
applicable to electric energy production in Germany and in other nations
also, having no access to huge storage devices. Second, this approach
leads to the prediction, that most of the present power demand of
Germany could be supplied by wind-solar power alone. Third, this approach
does no longer exclude the hope that even if most of the energy production
in Germany is switching to electric energy - which means a factor
of about $5$\citep{key-1} -, this energy can be delivered by wind-solar
power. In this case, however, no matter how we slice it, alone the
number of required wind turbines would become tremendous: $430000$
{[}110000{]} wind turbines of the $1.5MW${[}6MW{]} type (height 120m
{[}200m{]}). Controlling the volatility according to our approach
would increase these numbers further by $50-100$\% . The running
costs would rise by a factor $(1+\alpha),$ $\alpha$ being the strength
of the additional electric power production, defined in section 2.
According to our calculations reasonable values for $\alpha$ are
in the range $[0.5,1.0]$.

\end{document}